\documentclass[aps,prd,twocolumn,groupedaddress,showpacs,floatfix]{revtex4}
\usepackage{graphicx}
\usepackage{amsmath}
\begin{document}
\title
{\vspace{1.0cm}
Study of Proton Expansion in (p,2p) Quasielastic Scattering at Large 
Transverse Momentum}
\author{Alan S. Carroll}
\affiliation{ Collider-Accelerator Department, Brookhaven National Laboratory,
 Upton, NY 11973}
\date{February 24, 2003}

\begin{abstract}
 The measured nuclear transparencies in targets of Li, C, Al, Cu and Pb at incident momenta of 6, 10, 
and 12 GeV/c have been used to study the rate of proton expansion connected with (p,2p) quasielastic
scattering at large momentum transfer. Simple models with linear or quadratic 
expansion of the effective cross section fail to simultaneously fit the measured transparencies 
at all three momenta.  If only the 6 and 10 GeV/c transparencies are fitted, satisfactory 
representations can be obtained when the expansion distances for protons  at 6 GeV/c are greater than 
6.4 fm(linear) and 4.0 fm(quadratic).
These distances are greater than those suggested by most Expansion models except the quadratic 'naive
expansion' picture. However, the transparencies are well represented by the Nuclear Filtering
model with no explicit expansion. 
\end{abstract}
\maketitle
%
%
\section{Introduction}
Nuclear transparency is the experimental measure of  the ability of hadrons to penetrate nuclear matter. 
The measured quantity is
the ratio of the integrated quasielastic(q.e.) scattering cross section in a nucleus to that measured 
under the same kinematic conditions in a free elastic scattering.  For proton-proton(p,2p) scattering the
transparency, T, equation is,
\begin{equation}
 T=\frac{\frac{d\sigma}{dt}\text{[(p,2p) q.e. in nucleus]}}
   {Z\frac{d\sigma}{dt}\text{[(p,2p) free elastic]}}  
\label{A1}
\end{equation}
where Z equals the number of protons in the nucleus.
 Mueller and Brodsky suggested that the transparency would be increased compared to
a Glauber calculation whenever the hadrons involved 
had undergone a q.e. scattering at large momentum transfer \cite{MB},\cite{SB}.    
This was because the scaling 
laws of large angle scattering suggested that the valence quarks in the hadrons,  were in a point like 
configuration (plc) at the time of interaction. This concept is generally referred to a color transparency(CT)
since the QCD interaction is considerably
reduced by the near proximity of the quark color charges in the plc.  
Then for  high momenta, the hadrons would 
expand sufficiently slowly over distances compared to nuclear radii to produce an anomalously high 
transparency compared to that predicted by standard Glauber models. The transparency would 
approach 1.0 as the momentum was increased.

A series of measurements at the Alternating Gradient Synchrotron (AGS) 
of Brookhaven National Laboratory of (p,2p) q.e. interactions 
have consistently indicated significant 
changes in transparency with incident momentum for scattering 
near $90^o$ in the c.m. \cite{E834}, \cite{IM}, \cite{E850}.
  
Ralston, Pire and Jain view the q.e. process differently 
\cite{JPRREV},\cite{RP1}, \cite{RP2}, \cite{JR}.  
In their picture the nuclear 
medium strongly attenuates the large transverse portion of the pp scattering amplitude.  Nuclear Filtering
is a process which is constant throughout the nuclear volume, and results in a reduced, but 
non-expanding hadron size.  Detailed descriptions of their calculations can be found in Ref. 6 and
Ref. 9.

It should be noted that the  increasingly accurate (e,e'p) transparency measurements at 
SLAC and Jefferson Laboratory have shown no significant increase in  transparency with increasing $Q^2$ 
\cite{Jlab}, \cite{SLAC1}, \cite{SLAC2}.  The difference between the (p,2p) and (e,e'p) transparencies 
is likely to be a reflection of the greater complexity of the (p,2p) amplitudes.

	 An initial study by Heppelmann suggested that the 
transparencies of AGS Experiment E834 could be described by an 
effective attenuation cross section which was
smaller than the free space value and constant  in value \cite{SH}.  The measurements of AGS E834
for  (p,2p) q.e. scattering on nuclei ranging from Li to 
Pb provide a unique opportunity to measure this expansion directly in a way that has not
been done previously \cite{E834}.  The purpose of this paper is to make a quantitative 
comparisons of these two classes of models(Expansion and Nuclear Filtering) in as fair an analysis as 
possible with the existing data from these (p,2p) transparency experiments.
Other analysis have emphasized the energy dependence, but this analysis is centered on the
A dependence of the transparency at 6 and 10 GeV/c.  Expansion models have generally predicted expansion 
distances for these AGS experiments to be comparable to nuclear radii.

\section{Data}
	A series of measurements made at the Alternating Gradient Synchrotron (AGS) have 
determined nuclear transparencies for a number of different momenta and nuclei.  These 
measurements of (p,2p) q.e.  scattering in nuclei indicated effective cross sections for 
absorption in nuclei, which vary with incident momenta from 6 to 14 Gev/c, and are in general 
significantly less than the measured pN total cross sections. The Carbon transparencies as a 
function of incident momentum for the 1998 data from E850 by Leksanov, et al \cite{E850}
and the 1994 data from E850 by 
Mardor, et al \cite{IM},  and the 1987 data from E834 by Carroll, et al  \cite{E834} are shown in
Figure 1.  Also included are the 1987 Al data from E834 which has been scaled as $(27/12)^{1/3}$
to indicate the approximate consistency of these two nuclei.

\begin{figure}[hbt]
\includegraphics[width=8cm]{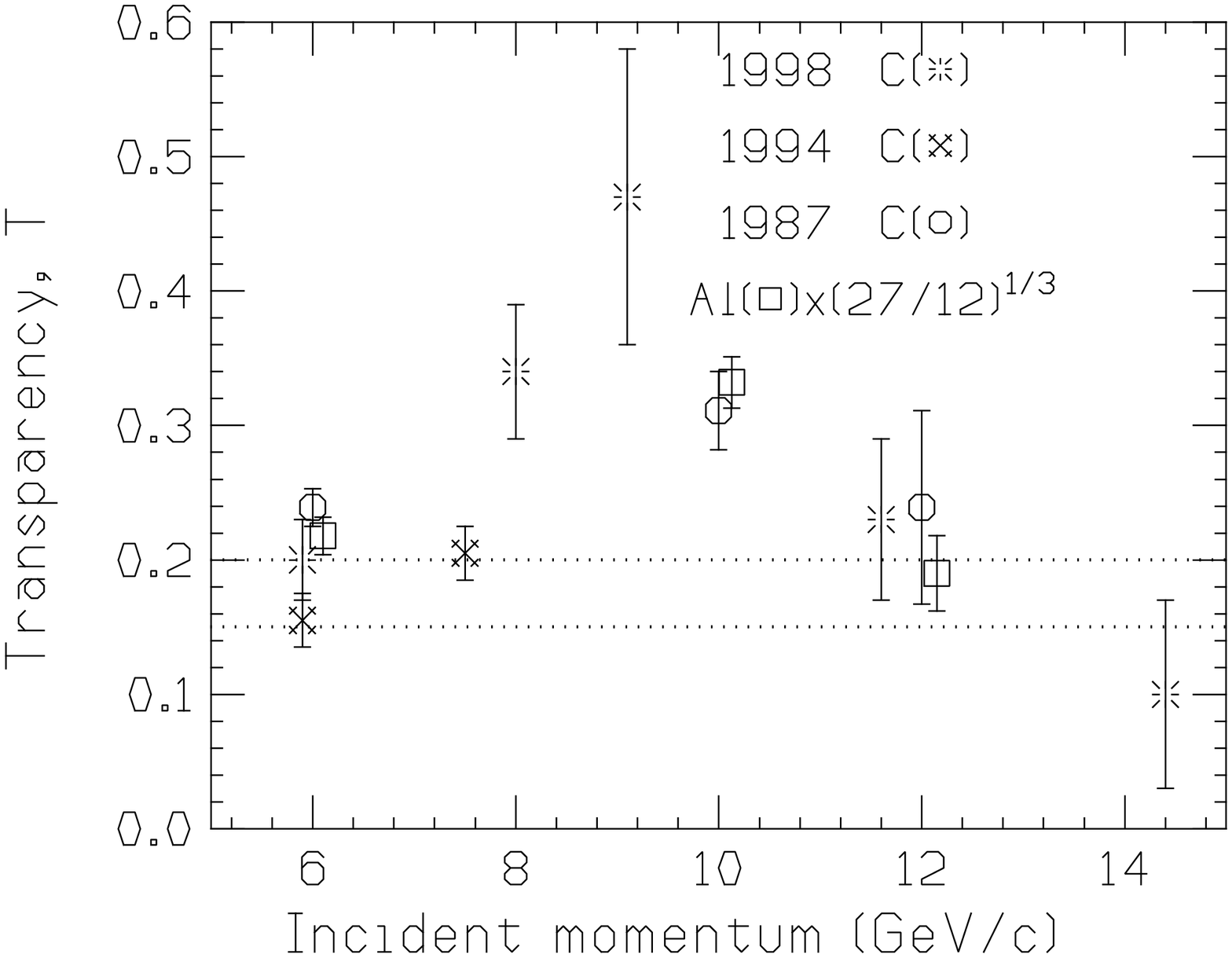}
\caption{\label{fig1}Summary of Nuclear Transparency $(p,2p)$ Measurements on Carbon and scaled
measurements on Aluminum(with small horizontal displacements for clarity). The dotted horizontal
lines indicate the range of Glauber calculations}
\end{figure}

	In particular, the publication of the data from the 1987 experiment 
reported on nuclear transparency values 
of Li, C, Al, Cu and Pb with natural isotopic abundances 
at incident momenta of 6 and 10 GeV/c, and C and Al at 12 GeV/c \cite{E834}.  
Measurements with all the targets indicated a clear increase in the transparency between 6 and 10 GeV/c.
At 12 GeV/c, the transparency of the C and Al nuclei
 was consistent with that at 6 GeV/c as shown in Fig 2.  Subsequent measurements 
with the new EVA spectrometer(E850) confirmed the transparencies for the C targets and expanded the 
range of momentum \cite{E850}.  

All the transparency values for the 5 nuclei  from E834
as plotted in Fig 1 and Fig 2 have been multiplied by a factor of 0.724.
This factor arises from the different methods for determining the transparencies in the two
experiments.  In E850, the transparency ratio was measured in a small region of the longitudinal light 
cone momentum, $\alpha=1.00\pm0.05$, corresponding to the struck proton being nearly at rest in the nucleus
\cite{IM},\cite{E850}.  Then the total transparency was calculated using a parameterization of the complete
Carbon spectral function \cite{CdA}.  For E834, the transparency was calculated using measurements of the
transparency from  essentially the 
entire range of $\alpha$.  The extraction of the transparencies involved a convolution with the 
spectral functions measured by the E834 experiment, and a correction for the energy dependence 
of the elementary pp differential cross section \cite{E834}.
The measured spectral functions used in the analysis of the E834 should correctly determine the relative
transparencies of the 5 nuclei, even though the procedure of E850 is felt to give a better absolute 
normalization.  Since the analysis of this paper includes a floating normalization, the change should
have little impact on the fitted results.

    Although the 12 GeV/c data are included for completeness, 
the fact that the 12 GeV/c transparencies are measured for only two 
adjacent nuclei with rather large errors means 
that the results are not strongly influenced unless all three momenta are tightly coupled through 
an Expansion picture.   

\begin{figure}
\includegraphics[width=8cm]{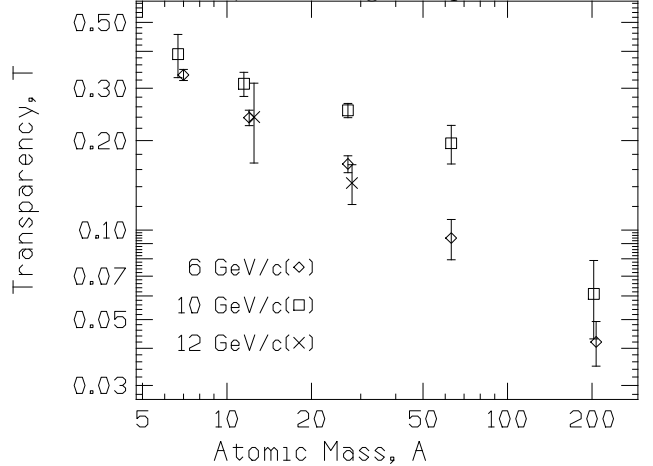}
\caption{\label{fig2} Transparency vs Atomic mass A for  $(p,2p)$ Measurements.  
}
\end{figure}

\section{Parameterization of expansion}

Two general classes of models have been developed to explain the behavior of hadronic interactions 
inside a nuclear medium.  In the Expansion class of models, the high $p_t$ interaction is presumed to select 
nearly point like configurations (plc's) of valence quarks in the 
the interacting protons.\cite{MB}  These plc's proceed to expand as their distance increases
 from the point of interaction.

	The second class of models emphasizes that in the nuclear medium, the major effect is to strongly 
attenuate the large transverse portion of the proton wave function.  This Nuclear Filtering
picture is primarily the work of Jain, Ralston, Pire \cite{RP1},\cite{JPRREV}. 
This model suggests that the effective cross section will be smaller than that of the free 
cross sections, and remain essentially constant as it passes through the nucleus.

The rate of expansion has been described in 
both partonic and hadronic representations \cite{FLFS},\cite{JM}. 
Farrar, Liu, Frankfurt, and Strikman suggested the 
expansion parameterization for the effective interaction cross section,
$\sigma_{eff}(z,Q^2)$ given by Eq 2 \cite{FLFS}. This form is a 
convenient one for this study:  
\begin{widetext}
\begin{equation}
\sigma_{eff}(z,Q^2)=
\sigma_{eff}^{\infty}\left(\left[\left(\frac{z}{l_h}\right)^{\tau}+
\left(\frac{r_{t}(Q^{2})^{2}}{r_{t}^{2}}\right)\left(1-\left(\frac{z}{l_h}\right)
^{\tau}\right)\right]\theta\left(l_h-z\right)+\theta\left(z-l_h\right)\right)
\label{eq:S}
\end{equation}
\end{widetext}

where $l_h$ is the expansion distance of the protons, and z is the distance from the interaction point.
 $\sigma_{eff}(z,Q^2)$  expands linearly or 
quadratically from its initial size depending on the value of $\tau$, and then assumes the
free space value,  $\sigma_{eff}^{\infty}$, when $z=l_h$.  As noted below, the actual value of 
$\sigma_{eff}^{\infty}$
used in the fitting procedure may be less than the free $\sigma_{tot}$(pN) for the proton-nucleon
interaction because a portion of the q.e. events with an initial or final state 
elastic scattering fall within the kinematical definition of a q.e event.   
Since all the measurements are made
near $90^o$ in the c.m., $Q^2=\sim p_0$.

The exponent $\tau$ allows 
for three suggested pictures of expansion; $\tau=0,1,\text{and}2$.
 For $\tau$ = 1, the expansion corresponds to the ``quantum diffusion'' 
picture \cite{FLFS}. For this picture, $l_h$ = $2p_f$/$\Delta$($M^2$)
where $p_f$ is the momentum of a  proton traveling through 
the nucleus and $\Delta$($M^2$) is the mass difference of an intermediate state
\cite{FLFS}.  At distances comparable to nuclear sizes, the 
effective cross sections should revert to their free space values. The authors of  \cite{FLFS} indicate
the values of   $\Delta$($M^2$)   between 0.5 and 1.1 $GeV^2$  are acceptable with  $\Delta$($M^2$)=0.7
being favored.  This range of 
 $\Delta$($M^2$)  corresponds to 
values of  $l_h$ =  0.36$p_f$ to  0.78$p_f$ fm.  For a momentum of 6 GeV/c the expansion distance will be 
between 2.1 and 4.7 fm.  .  For convenience of calculation in this paper an 
expansion parameter, $\lambda$,  
scaled to 6 GeV/c has been used to parameterize all the proton momenta in the interaction for
each incident momentum.  That is the expansion distance $l_h$ for each leg of the calculation
shown in Fig 3 is given by $l_h$= $\lambda(p_{f}/6)$ fm.

	The case of $\tau=2$ is generally referred to as the 'naive quark expansion' 
scenario in which the light
quarks fly apart at a maximum rate and the distance is determined by the Lorentz boost to the hadrons.
In this case $l_h=\sim{E/m_h}$ where $m_h$ is the mass of the hadron
involved \cite{FLFS}. For protons at 6 GeV/c, $\lambda$ equals $\sim7.3$ fm.

  The quantity  $<r_{t}(Q^{2})^{2}>/<r_{t}^{2}>$ represents the  fraction of $\sigma_{eff}$ at 
the time of interaction.  
This quantity is approximated by $\sim{1/Q^2}$, corresponding to 0.21 at 6 Gev/c and 
falling  with an increase of incident momentum \cite{FLFS}.  Variations 
of this value have only a small  effect on the result.  A recent analysis by Yaron, et al 
repeats this analysis with $\tau=1$ and obtains very similar results \cite{YFP}.

Given that the initial and final states in these (p,2p) q.e. interactions are exclusive
hadrons, the approach of Jennings and Miller to represent the proton expansion in terms of
a set of hadronic states seems very reasonable  \cite{JM},\cite{JM2}.  This representation explicitly
notes that a plc cannot be a simple proton, but must include a superposition of excited states.  When
this spectrum of intermediate states,  $g(M^{2}_{X})$, is described by a power law falloff, then
the expansion has a linear form, $\tau=1$, with $\lambda=\sim0.9$ fm \cite{JM}.  With a sharp cutoff
of $M_X^2$ at $\sim2.2GeV^2$, then $\sigma_{eff}$  grows quadratically with $\lambda=\sim2.4$ fm \cite{JM}.
The form of these expansions can be approximated by that given in Eq. 2.


 Because in the Nuclear Filtering picture, the long distance portion of the amplitude has been 
filtered away by the nuclear medium, the cross section for q.e. scattering 
in the nucleus will follow the scaling behavior, whereas the unfiltered free pp cross section will show 
oscillations about the $s^{-10}$ scaling. 
 Thus the variations in the nuclear transparency are mainly due to 
the oscillations in the free pp cross section about the cross section with exact scaling.
 In fitting the transparency, no expansion should be 
required, only a smaller effective cross section.  In this model of the second class, 
$\tau$  is set to 0, so Eq. 2 reduces to, 

\begin{equation}
\sigma_{eff}(z,Q^2)=\sigma_{eff}(Q^2).
\end{equation}  

Also $\sigma_{eff}(Q^2)$ is allowed  to vary over an extended range of values. 
This analysis is very similiar to that described by Jain and Ralston \cite{PJ}.

\section{Method}

	There seems to be no simple parameterization of nuclear transparency as a function of incident 
momentum ($p_0$), nucleus(A), effective cross section ($\sigma_{eff}$) and expansion distance ($\lambda$).
 So the approach 
taken in this paper is to calculate via Monte Carlo means the nuclear transparency at a number of 
closely spaced values, and then do a search to find the best fit to the experimental values.

\begin{figure}
\includegraphics[width=8cm]{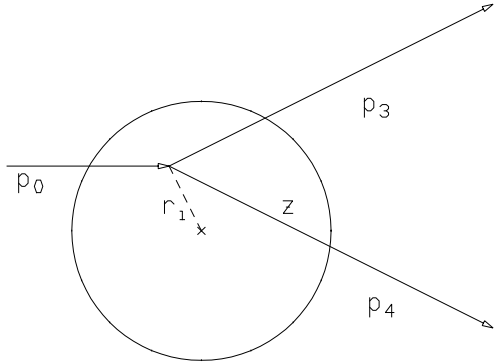}
\caption{\label{Fig_3} Coordinates for Transparency Calculations by Monte Carlo.} 
\end{figure}

  Fig 3 illustrates the geometry and kinematics of these calculations.
The integrals for the calculating the transparency at each 
incident momentum, $p_0$ and outgoing momentum, $p_3$ and
$p_4$,  are given by the following expressions. Also a normalization parameter, $r_n(p_0)$,
as described below, is included.

\begin{equation}
     T(\sigma_{eff},A,\lambda)=r_n(p_0)P_0P_3P_4
\end{equation}

where the average survival probabilities, $P_i$, of the protons on each of the three legs (i) is calculated 
by the integrals along each of the three paths in $z$ from the 
from randomly selected interaction points to the edge of the nucleus,

\begin{equation}
    P_i=exp[-\int_{path}dz'\sigma_{eff}(p_i,z,\lambda_{i})\rho_A(r_i)]
\end{equation}	

A Woods-Saxon form was used for the density,  $\rho(r_i)=c/(1+exp(-R+r_i/b))$, where $r_i$ is the
radial distance from the nucleus center to a point along the $i^{th}$ path. The 
parameter b is set to 0.56 fm, and 
then the $<rms>$ radii were matched to electron scattering results \cite{DEV}. The
integrated density was normalized to be equal to the A of the nucleus.
 The calculated Carbon transparencies are in agreement with Glauber calculations
of the Carbon transparency of 0.15 to 0.20 for  $\sigma_{eff}$ between 36 and 40 mb by a number of 
other authors \cite{MZ1},\cite{MZ2},\cite{ML}. 
Since this approach is deeply related to the Glauber theory similar results from that 
method can be expected.  The Expansion model uses a classical cross section, and it is
best that it be be tested on the same basis as it was formulated.
Note that this agreement with the Glauber calculation for 
the Carbon nucleus alone does not imply agreement with a A dependence for all 5 nuclei.
The effect of correlations on the calculated transparencies and the fitted results is
discussed in the Appendix following Lee and Miller \cite{ML}.  

The transparency was calculated for the 5 different nuclear targets in 1 mb steps of 
$\sigma_{eff}$  from 1 to 51 mb, and at 14 values of $\lambda$ from 0 to 50 fm. Then  the 
calculated transparencies at each value of A, $p_0$, and $\sigma_{eff}$ are 
parameterized with an empirical four parameter function in $\lambda$,
$T(\lambda)=\alpha+(1-\alpha)e^{[\beta/(zc+\lambda)]}-\gamma\lambda$,
 for use in fitting to the measured transparencies.  Note 
that the  $\lambda$ value couples the expansion between 6, 10, and 12 GeV/c. The 1000 trials generated for 
each point resulted in a statistical accuracy of $\pm0.01$ in the calculated transparency values.  
As an illustration, a sample of the calculated transparency values for 10 GeV/c is in given in Table I. 

Using the generated values of the transparency, a best fit was made to the values for 6, 10, 
and 12 GeV/c.  The random search was made by interpolating between 
 $\sigma_{eff}$  values for the 5 
different nuclei, and calculating the fitted $T(\lambda)$ function . The search determined 
the best fit from minimizing the $\chi^2$ function given in Eq. 6.

\begin{widetext}
\begin{equation}
   \chi^2=\Sigma_{i=1}^{5}S^{2}_{i}(6 GeV/c)+\Sigma_{i=1}^{5}S^{2}_{i}(10 GeV/c)+
\Sigma_{i=2}^{3}S^{2}_{i}(12 GeV/c)
 \end{equation}  
\end{widetext}

 where there are sums, $\Sigma$, for the three momenta and the 5 nuclei (2 nuclei at 12 GeV/c).
 The terms for each momentum (k) and nucleus(i) are of the form:
\begin{equation}
     S^{2}_{i}(p_k)=[(r_n(p_{0,k})T_{i}(fit)-T_{i}(meas))/( \Delta T_{i}(meas))]^2                             
\end{equation}

\begin{table}
\caption{\label{tab:sample}
 Sample of calculated transparency values for 10 GeV/c with an initial 
$\sigma_{eff}$ = 0.117$\sigma_{eff}^{infty}$.
}
\begin{ruledtabular}
 \begin{tabular}{|l|c|c|c|r|}
      A  &    $p_0$ &   $\sigma_{eff}$ & $\lambda$ &   Transp  \\
         &    GeV/c &   mb &  fm &  T     \\
\hline
     7.0   &   10.0  &   1.00  &  0.00  &  0.9603   \\
     7.0   &   10.0  &   1.00  &  0.50  &  0.9636   \\    
      .    &     .   &    .    &   .    &   .       \\
      .    &     .   &    .    &   .    &   .       \\
      .    &     .   &    .    &   .    &   .       \\
    12.0   &   10.0  &  19.00  &  13.50 &  0.7645   \\
    12.0   &   10.0  &  19.00  &  16.50 &  0.7811   \\ 
    12.0   &   10.0  &  19.00  &  50.00 &  0.8387   \\
      .    &     .   &    .    &    .   &   .       \\
    27.0   &   10.0  &  19.00  &   0.00 &  0.2373   \\
    27.0   &   10.0  &  19.00  &   0.50 &  0.2542   \\
    27.0   &   10.0  &  19.00  &   1.00 &  0.2838   \\
    27.0   &   10.0  &  19.00  &   1.50 &  0.3227   \\
    27.0   &   10.0  &  19.00  &   2.50 &  0.3597   \\
    27.0   &   10.0  &  19.00  &   3.50 &  0.4194   \\
    27.0   &   10.0  &  19.00  &   4.50 &  0.4652   \\
    27.0   &   10.0  &  19.00  &   5.50 &  0.5126   \\
    27.0   &   10.0  &  19.00  &   6.50 &  0.5401   \\
    27.0   &   10.0  &  19.00  &   7.50 &  0.5750   \\
    27.0   &   10.0  &  19.00  &  10.50 &  0.6156   \\
    27.0   &   10.0  &  19.00  &  13.50 &  0.6548   \\
    27.0   &   10.0  &  19.00  &  16.50 &  0.6850   \\
    27.0   &   10.0  &  19.00  &  50.00 &  0.7691   \\
      .    &     .   &    .    &    .   &   .       \\
    63.5   &   10.0  &  19.00  &   0.00 &  0.1368   \\
    63.5   &   10.0  &  19.00  &   0.50 &  0.1624   \\
      .    &     .   &    .    &    .   &   .       \\
      .    &     .   &    .    &    .   &   .       \\
      .    &     .   &    .    &    .   &   .       \\
   207.2   &   10.0  &  51.00  &  16.50 &  0.1011   \\
   207.2   &   10.0  &  51.00  &  50.00 &  0.2116   \\
\end{tabular}
\end{ruledtabular}
\end{table}

Note that in addition to the values in the table of generated transparencies, relative normalization 
factors for each incident momentum,  $r_n(p_{0,k})$ , are included to allow for normalization 
uncertainties in both the data and the uncertainties of the phenomenological 
transparency calculations for each incident momentum (k). A similar factor was used in the analysis
of Jain and Ralston \cite{PJ}. The search procedure used for each value of 
$\lambda$ in steps of 1.0 from 0 to 20 was straightforward.  Values for  $\sigma_{eff}$ 
 and $r_n(p_{0,k})$ were randomly selected for the  entire 
range of possible values, and then the values which yielded the smallest fitted $\chi^2$  were selected.
For the Expansion models, the values $\sigma_{eff}$ at 6, 10 and 12 Gev/c are constrained to be
equal, and to the values of $r_n(p_{0,k})$  are allowed to vary by up to $\pm 15\%$ with respect to each other
to allow for relative normalizations.  The fitting procedure was applied to both the sets of 
transparencies at 6, 10 and 12 GeV/c, and the set containing only 6 and 10 GeV/c transparencies.
A $4{\times}10^5$ trial coarse search over the full range of variables  was followed by a 
$6{\times}10^5$ trial fine search within 10\% of the final values.  Repeated 
applications of this procedure yielded fits, which varied by at most 1\%.  The quality of the fits is 
indicated by the value of $\chi^2$ .

	The Expansion models assume that  $\sigma_{eff}$  returns to its free space values at
 some distance.  
At the incident momenta of this experiment, that distance is expected to be comparable to the radius 
of the heavier nuclei.  Not all of the elastic scattering cross section ($\sim{8mb}$ out of 40 mb)
 should  be included in  $\sigma_{eff}$ because the kinematics of some of the q.e.
 events with initial and final state elastic scattering reconstruct within 
the Fermi distribution of $\sim250$ MeV/c. 
  A Monte Carlo study of the experimental acceptance
of E834 indicates that only  $2.5 \pm 1.0$ mb of the elastic cross section  
should be included.  For purposes of this study, all $\sigma_{eff}$  values above 32 mb have been 
allowed in the fitting procedure for the Expansion models so that the result is not dependent on the precise 
magnitude of the elastic cross section 
included.  The maximum  $\sigma_{eff}$  allowed, 45 mb, is well beyond the maximum
expected.  

	For the fit with the Nuclear Filtering model, the parameter, 
$\tau$ is set to 0, and  $\sigma_{eff}$ is allowed to 
vary from 1 to 45 mb at each incident momenta.
\section{Results}

	Fig. 4 gives the result of fitting the transparencies to the linear ($\tau=1$) Expansion hypothesis.
As stated above, the values of $\sigma_{eff}$ are constrained to be greater than 32 mb, and equal in magnitude at
each step in $\lambda$.  The values of $r_n(p_{0,k})$ are held to be within $\pm 15\%$ of each other at each step.
The solid curve starting at $\sim{60}$ corresponds to $\chi^2$ in the fit to the 6, 10 and 12 GeV/c transparencies.
The minimum value of this  $\chi^2$ curve is 19 which has a probability of 1.5\%.  The upper dot-dash curve gives the
fitted value for $\sigma_{eff}$ which stays at the minimum value of 32 mb for $\lambda$ $<$ 6 fm.
The lower dot-dash line corresponds to  $r_n(p_{0,k})$ multiplied by 10.  $r_n(p_{0,k})$ falls from a value
of $\sim{1.0}$ at $\lambda$=0 fm to 0.5 at larger expansion distances. The values of  $r_n(p_{0,k})$ are held
to within $15\%$ of one another.

\begin{figure}
\includegraphics[width=8cm]{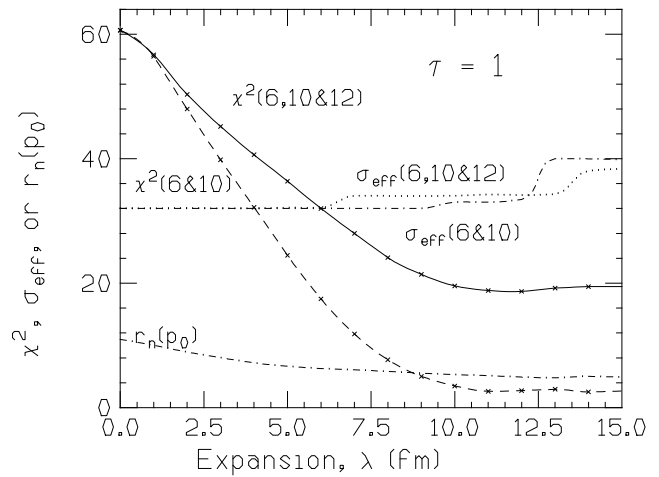}
\caption{\label{fig4} The $\chi^2$ and fitted parameters for a fit to 
transparencies with linear Expansion model ($\tau=1$). $\sigma_{eff}$ is given in mb.
The normalization parameter, $r_n(p_{0})$, have been multiplied by 10.   }
\end{figure}

	The dashed curve in Fig 4 starting at $\sim{60}$ is the $\chi^2$ for fitting 
only the 6 and 10 GeV/c transparencies.
The probability of $\chi^2$ reaches 5\% for values of $\lambda$ greater than 6.4 fm. The dotted curve traces 
the behavior of  $\sigma_{eff}$ for this fit.  The values of  $r_n(p_{0,k})$ are very similar in both cases.

\begin{figure}
\includegraphics[width=8cm]{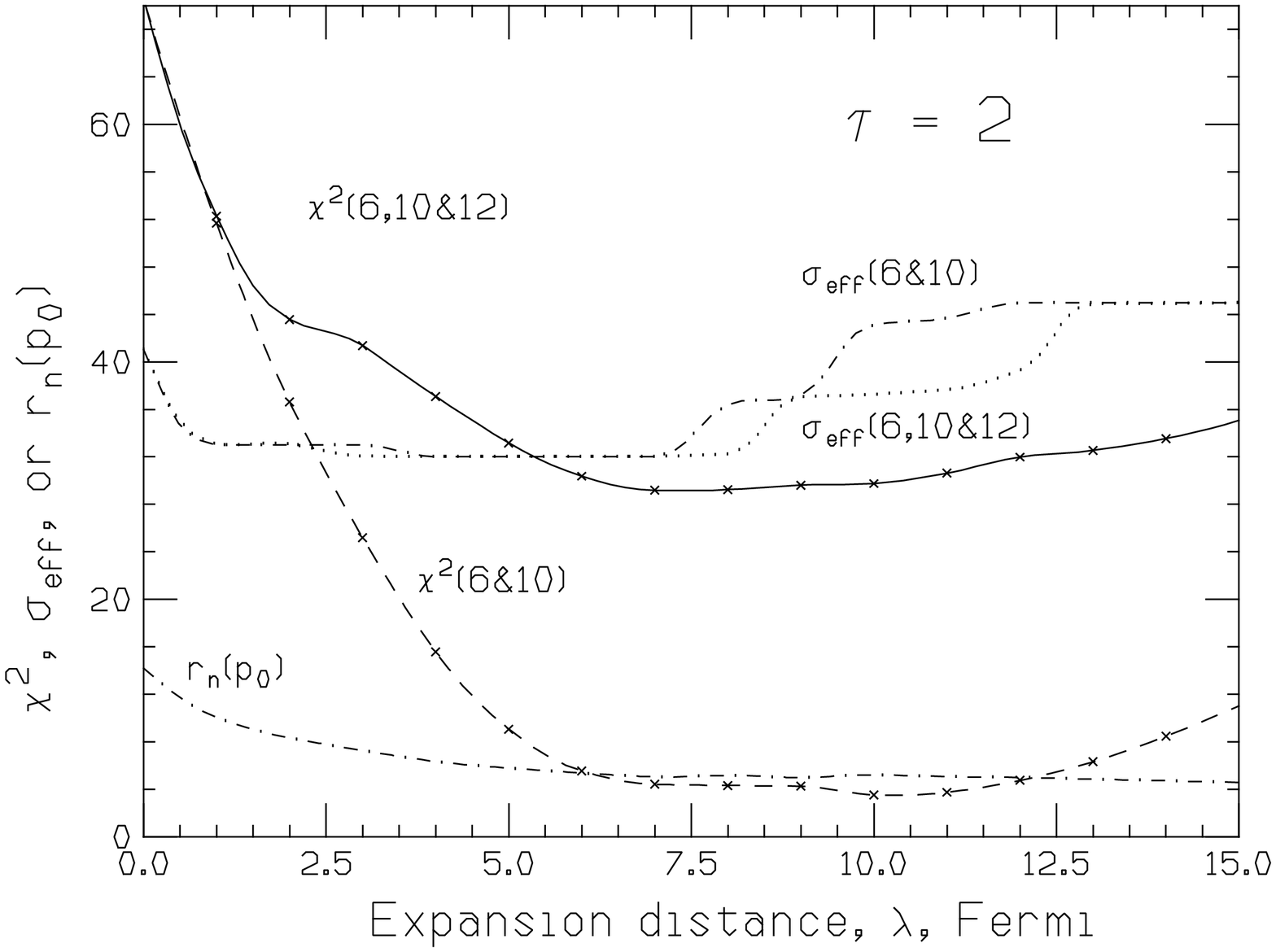}
\caption{\label{fig5}Fit to transparencies with quadratic Expansion model ($\tau=2$).  The meaning of the
curves is the same as in Fig 4.}
\end{figure}

	The results of fitting to the quadratic expansion ($\tau=2$) are shown in Fig 5.  The curves have the same 
meaning as in Fig 4.   $\chi^2$ for the fit to the 6, 10 and 12 GeV/c transparencies (solid curve) never goes below 29.2,
 corresponding to a probability of less than 0.012.  For the case of a fit to only the 6 and 
10 GeV/c data (dashed curve), the probability reaches 5\% at $\lambda$=4.0 fm.

\begin{table}
\caption{\label{tab:results}
 Parameters for Nuclear Filtering ($\tau=0$) case with 8 DoF. With no constraint on $r_n(12 GeV/c)$ the 
value of $\sigma_{eff}(12 GeV/c)$=$19^{+21}_{-15}$ mb. 
}
\begin{tabular}{|l|r|}
\hline
\hline
$r_n(6 GeV/c)$ & $0.63\pm0.02$  \\
\hline
$\sigma_{eff}(6 GeV/c)$ (mb) & $17.9^{+2.7}_{-1.5}$ \\
\hline
\hline
$r_n(10 GeV/c)$ & $0.65\pm0.02$  \\
\hline
$\sigma_{eff}(10 GeV/c)$ (mb) & $12.3^{+2.6}_{-2.6}$ \\
\hline
\hline
$r_n(12 GeV/c)$ & $0.59\pm0.02$  \\
\hline
$\sigma_{eff}(12 GeV/c)$ (mb) & $19.0^{+8.9}_{-3.5}$ \\
\hline
\hline
$\chi^2$ & $3.77$  \\
\hline
$Prob(\chi^2)$ & $87\%$ \\
\hline
\hline
\end{tabular}
\end{table}

	Table II displays the values of a fit with Nuclear Filtering ($\tau=0$). 
  Here the values of $\sigma_{eff}$ are allowed to vary 
independently at each momenta without constraints on the minimum value of  $\sigma_{eff}$.  However, 
the values of  $r_n(p_{0,k})$ are again constrained to remain within $\pm15\%$ of each other. The overall 
$\chi^2$ of 3.77 indicates a probability of 87\% for 8 DoF. The errors are determined from
the one standard deviation in the ln(Likelihood) \cite{PDG}.  Jain and Ralston found values of
$17\pm2$ mb and $12\pm2$ mb for $\sigma_{eff}(6 GeV/c)$
and $\sigma_{eff}(10 GeV/c)$ which are consistent those in Table II \cite{PJ}.

\begin{figure}
\includegraphics[width=8cm]{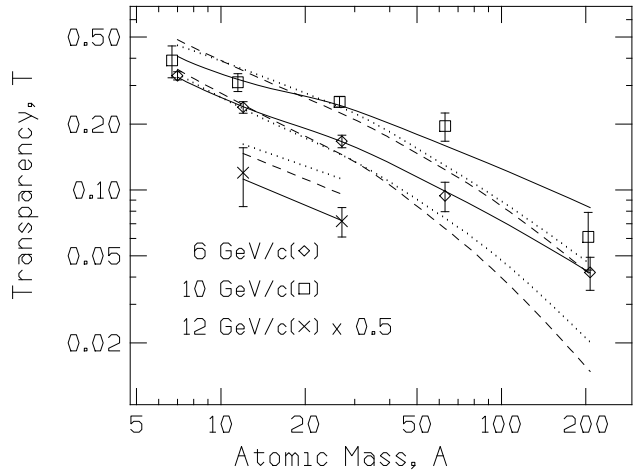}
\caption{\label{fig6} Representative fits to transparencies. The Nuclear Filtering model $\tau=0$ 
is represented by the solid curves, and the $\tau=1$ and $\tau=2$ Expansion models at $\lambda=3$ fm
 are displayed as the dashed and dotted curves respectively. Note that the 12 GeV/c transparencies have
been multiplied by 0.5 to avoid overlap with the 6 GeV/c results.}
\end{figure}

	Fig. 6 illustrates the quality of the fit to the experimentally measured
transparencies for each of the 5 nuclei at 6, 10,
 and 12 GeV/c for the 3 models; namely for  $\tau=0$ for the Nuclear Filtering model, and 
 $\lambda$=3 fm for $\tau=1$ and $\tau=2$. At this expansion distance, the $\tau=1$ and $\tau=2$ 
Expansion models
indicate a fall off of transparency with A which is much steeper than that measured.
 Generally reasonable fits can be made with the $\tau=1$ and $\tau=2$ expansion models  to the 6 and 10 GeV/c
transparencies alone when
of $\lambda$ is greater than 6 fm. However, 
only the  Nuclear Filtering (solid curve) can simultaneously fit to the 6, 10 and  12 GeV/c transparencies.

\section{Conclusions}

	Table III presents a summary of this analysis, and predictions of various models.
Due to the oscillatory nature of  the (p,2p) transparency with incident momentum, it is 
not surprising that no acceptable fit with $Prob(\chi^2) > 0.05$ can be achieved with a simple, 
unified Expansion model simultaneously 
fitting to the data at 6, 10 and 12 GeV/c.  As has been noted by various authors, additional amplitudes are
needed  to account for the sudden drop in transparency between 10 and 12 GeV/c.
This measured drop in the
transparency has been verified by the E850 experiment, and is shown in Fig 1 to continue 
to higher momenta \cite{E850}.  

For Ralston and Pire, the drop in transparency is connected with the interference
of the short distance pQCD amplitude with that of the long distance Landshoff contribution  \cite{RP1}. 
Brodsky and deTeramond  \cite{BdeT}  noted the strong correlation in energy between the
striking spin dependence of pp scattering \cite{AK} and the behavior of the (p,2p) transparency \cite{E834}.
They suggested that the drop in transparency at 12 GeV/c could be due to the presence
of a resonance in the pp channel creating a long-range amplitude.  This resonance
could be connected with the threshold of charm particle production \cite{BdeT}.


\begin{figure}
\includegraphics[width=8cm]{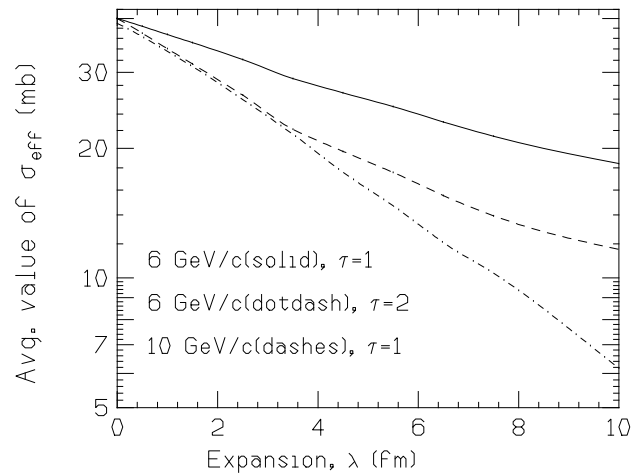}
\caption{\label{fig7} Average $\sigma_{eff}$ vs $\lambda$ for Aluminum .}
\end{figure}

	One might imagine that the 6 and 10 GeV/c transparencies represent a simpler set of data
where only one set of amplitudes dominate.  Thus a simple Expansion hypothesis could be satisfied.
This is the motivation for showing how the Expansion models fit the 6 and 10 GeV/c data alone.

\begin{widetext}
\begin{table*}
\begin{ruledtabular}
\caption{\label{tab:summary}Summary of Fit Parameters and Comparison to Models}
\begin{tabular}{|l|c|c|c|c|c|}
\hline
This Analysis & $\tau=0$ & $\tau=1$ & $\tau=1$ & $\tau=2$ & $\tau=2$ \\
Momenta Fit   & $6,10\&12$ & $6\&10$ & $6,10\&12$ & $6\&10$ & $6,10\&12$ \\
\hline
Prob($\chi^2$) & $0.87$ & $>0.05$ & $<0.044$ & $>0.05$ & $<0.012$ \\
  .            & const. & for    &   for     &  for   &   for    \\   
$\lambda$,fm   &        & $>6.4$   &   All     &  $>4.0$ &   All    \\
$\sigma_{eff}(6GeV/c)$, mb & $17.9^{+2.7}_{-1.5}$ & & & & \\
$\sigma_{eff}(10GeV/c)$, mb &  $12.3^{+2.6}_{-2.6}$ & & & & \\
\hline
\hline
Farrar, et al & & &  & & \\
Prob($\chi^2$) & & $1\times10^{-7} - 8\times10^{-4}$ & $3\times10^{-7} - 5\times10^{-5}$ & $0.82$  & $1\times10^{-3}$  \\
$\lambda$,fm  & & 2.1 - 4.7 & 2.1 - 4.7  & 7.3 & 7.3 \\
\hline
Jennings - Miller & & & & & \\
Prob($\chi^2$) & & $2\times10^{-9}$ & $1\times10^{-8}$ & $1\times10^{-4}$  & $6\times10^{-6}$  \\
$\lambda$,fm  & & 0.9 & 0.9  & 2.4 & 2.4 \\
\hline
\end{tabular}
\end{ruledtabular}
\end{table*}
\end{widetext}

	Since there are values of  $\lambda$ for which the 6 and 10 GeV/c data alone can be satisfactorily
fit, it is interesting  to consider whether these values of  $\lambda$ agree with various models.
The maximum value of expected value of $\lambda$ for the  linear ($\tau=1$) expansion corresponds to an
intermediate mass, $\Delta(M^2)$=0.5 $Gev^2$, corresponding to 
 $\lambda$=4.7 fm at at 6 GeV/c \cite{FLFS}.  At this value Prob($\chi^2$) is  $8\times10^{-4}$.
The hadronic model suggests that  $\lambda$=0.9 fm  \cite{JM},\cite{JM2}.
Thus no linear Expansion pictures in either the partonic or hadronic representations provide expansions
long enough to fit the data.

The curves of Fig 7 show a calculation of the average value of  $\sigma_{eff}$ over the path lengths 
in the Al nucleus for a range of expansion parameters,  $\lambda$. At expansion distances of $\sim{6}$ fm, the
average  $\sigma_{eff}$ approaches the fitted values of  $\sigma_{eff}$ in the 
$\tau=0$  case indicating how the large values of $\lambda$ yield acceptable fits in the
case of expansion.

 	For the $\tau=2$ expansion, an acceptable fit to 6 and 10 GeV/c 
is reached at a smaller value of  $\lambda$ 
 due to the more rapid fall off of  $\sigma_{eff}$ with $\lambda$ (see Fig 7).
The Prob($\chi^2$) becomes 5\% at 4.0 fm  which is
 within the range of   $\lambda$= 7.3 fm suggested by the 'naive Expansion model' \cite{FLFS}.
In the hadronic representation of Jennings and Miller, a quadratic expansion has a 
$\lambda$ of $2.4$ fm \cite{JM} which has a probability of $1\times10^{-4}$.

	The Nuclear Filtering picture is favored by this analysis.  There is a different constant value of 
 $\sigma_{eff}$ each incident momentum, and hence $Q^2$. 
However,  $\sigma_{eff}$ shows no expansion over range of nuclear radii from 
Li (2.1 fm) to Pb (6.6 fm) and  provides an acceptable description of the data as has been shown in 
previous publications\cite{SH},\cite{JR}.
Both linear and quadratic expansion pictures fail to fit the entire set
of data.  Fits to the limited 6 and 10 GeV/c data set are achieved for linear expansions which are 
beyond the range of a variety of models.  The quadratic expansion ($\lambda=\sim{7.3} fm$)
in the naive quark picture
can provide an acceptable fit to
the 6 and 10 GeV/c data, but the theoretical basis for such simple behavior seems weak. 
However, as indicated in Fig. 7, the quadratic fit confirms the need  for a small  $\sigma_{eff}$.

	For future (p,2p) experiments it would be very interesting to measure the A dependence for 
an incident momentum in the range of 12 to 14 GeV/c where the transparency is at a minimum.  According to
the Jain, Pire and Ralston picture\cite{JR}, the  value of  $\sigma_{eff}$ should continue
to decrease even though the transparency has fallen by about of factor of two from its C value at
9 GeV/c.
\begin{acknowledgments}
I am pleased to acknowledge the dedicated efforts of all my collaborators on 
E834\cite{E834} and E850\cite{IM},\cite{E850}.
In particular, I would like to express my appreciation to my co-spokesman, S. Heppelmann;
my Brookhaven colleagues, D. Barton, G. Bunce, and Y Makdisi; H. Nicholson of 
Mt Holyoke College; and my collaborators from 
Tel Aviv, J. Alster and E. Piasetzky.  The helpful discussions with  P. Jain and J. Ralston
were greatly appreciated. Improvements to the text were generated by the careful
proof reading of Y. Makdisi.
\end{acknowledgments}
\appendix*
\section{Effect of Correlations}
The Monte Carlo calculations of transparencies were adjusted to match existing Glauber
calculations of C transparency.  Explicit correlation effects were not included
in the calculations of this paper.  See Ref \cite{FMSS} for a discussion of correlation
effects. As a check, some of the calculations were repeated using 
the formulation of Lee and Miller \cite{ML}.
 These correlations indicate the nuclear density seen by the
outgoing protons is reduced for a distance of $\sim1 fm$ in the vicinity of the struck proton. 
A comparison of the results for the linear expansion($\tau=1$) with and without the
correlation correction for the 6 and 10 GeV/c fit is shown in the Fig 8.

\begin{figure}
\includegraphics[width=8cm]{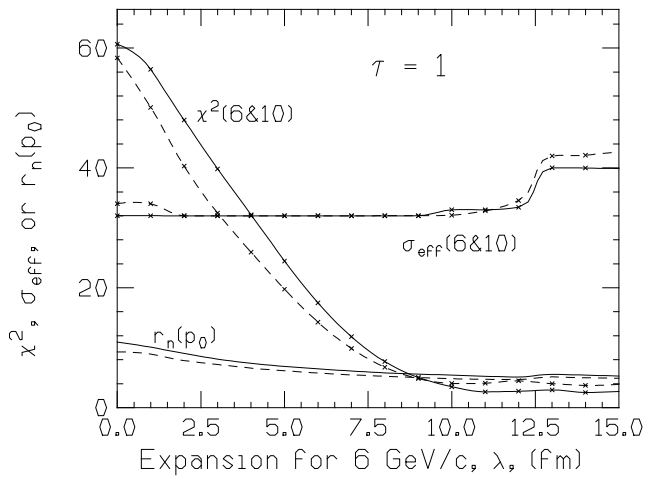}
\caption{\label{fig8} Result of fitting with Correlations (solid lines) and without Correlations(dashed lines)}
\end{figure}

As can be seen there is little difference in the parameters or the quality of the fit.
The correlations increase the transparencies by $\sim0.05$ for C and $\sim0.005$ for
Pb at $\lambda=5 fm$.  A small adjustment (0.688 to 0.616 for $\lambda=5fm$) of the normalization 
parameter,  $r_n(p_{0,k})$, suffices to achieve nearly the same $\chi^2$.  

Thus the conclusion is reached that the results are not very sensitive to the exact form of the nuclear density.


\end{document}